\documentclass{emulateapj}

\lefthead{A. van der Wel}
\righthead{Galaxy Morphology and Structure}
\slugcomment{The Astrophysical Journal, 675: L000-L000, 2008, March 1}

\begin{document}

\title{The Dependence of Galaxy Morphology and Structure\\ 
on Environment and Stellar Mass}

\author{Arjen van der Wel\altaffilmark{1}}

\altaffiltext{1}{Department of Physics and Astronomy,
Johns Hopkins University, 3400 North Charles Street, Baltimore, MD
21218; e-mail: wel@pha.jhu.edu}

\begin{abstract}
From the Sloan Digital Sky Survey (SDSS) Data Release 5 (DR5), we
extract a sample of 4594 galaxies at redshifts $0.02<z<0.03$, complete
down to a stellar mass of $M=10^{10}~M_{\odot}$. We quantify their
structure (S\'ersic index), morphology (S\'ersic index +
``Bumpiness''), and local environment.  We show that morphology and
structure are intrinsically different galaxy properties, and we
demonstrate that this is a physically relevant distinction by showing
that these properties depend differently on galaxy mass and
environment.  Structure mainly depends on galaxy mass whereas
morphology mainly depends on environment. This is driven by variations
in star formation activity, as traced by color, which only weakly
affects the structure of a galaxy but strongly affects its
morphological appearance.  The implication of our results is that the
existence of the morphology-density relation is intrinsic and not just
due to a combination of more fundamental, underlying relations. Our
findings have consequences for high-redshift studies, which often use
some measure of structure as a proxy for morphology. A direct
comparison with local samples selected through visually classified
morphologies may lead to biases in the inferred evolution of the
morphological mix of the galaxy population, and misinterpretations in
terms of how galaxy evolution depends on mass and environment.
\end{abstract}
\keywords{galaxies: fundamental parameters---galaxies: statistics---galaxies: structure}

\section{INTRODUCTION}

The morphology-density relation \citep[MDR,][]{dressler80} implies
that the environment affects the star formation history, color, and
structure of galaxies.  Many possible mechanisms have been suggested
to explain the suppressed or quenched star formation activity of
galaxies located in dense environments, most notably ram pressure
stripping \citep{gunn72}, harassment \citep{farouki81,moore96},
strangulation \citep[e.g.,][]{larson80,kauffmann93,diaferio01}, and
tidal interactions and merging \citep[e.g.,][]{park07b}.  At the same
time, it is now also well established that strong correlations exist
between galaxy mass and, for example, color \citep[e.g.,][]{baldry06}
and star-formation rate \citep[e.g.,][]{brinchmann04}. Several
suggestions have been made to explain the dependence of star formation
history on galaxy mass, all of which are related to feedback
mechanisms via either supernovae \citep[e.g.,][]{white91}, AGNs
\citep[e.g.,][]{croton06}, or shock heating of infalling gas
\citep{dekel06}.

Galaxy structure as measured by, e.g., concentration or S\'ersic
index, behaves differently from color and star formation in the sense
that structure depends strongly on galaxy mass but only weakly on
environment \citep{hogg04,kauffmann04}.  In the context of the MDR
this may be surprising as it is clear that concentration and
morphology are closely related quantities \citep[e.g.,][]{bell03}.
One possibility is that morphology and concentration only correlate
with environment through the underlying correlation between
environment and galaxy mass.  Another possibility is that morphology
and concentration are intrinsically different galaxy properties. For
example, it has been suggested \citep[by, e.g.,][]{kauffmann04} that
morphology is strongly related to star formation, more so than
concentration.

\citet{blakeslee06} developed an automated, quantitative scheme, the
$B$-$n$ method ($B$ for ``Bumpiness'', and $n$ for \citet{sersic68}
index, see Sec. 2), to distinguish E+S0 galaxies from later types in
\textit{Hubble Space Telescope} imaging of distant clusters.  In
\citet{vanderwel07b} (hereafter, vdW07) we generalize the $B$-$n$
method to classify galaxies both at high redshift and in the Sloan
Digital Sky Survey (SDSS).  In both cases the classifications from the
$B$-$n$ method agree very well with visually determined morphologies
as long as only two types are considered. In the current Letter we
study, for a local sample extracted from the SDSS, the relation
between structure (in this Letter measured by S\'ersic index $n$),
morphology (measured by $n$ and $B$), color, mass, and environment.
This allows us to decide which factors determine the morphological
appearance of a galaxy, and provides insight into whether or not the
mechanisms that are responsible for morphological transformations are
identical to the processes that suppress star formation.

We use the Fifth Data Release (DR5) from the SDSS \citep{adelman07}
and adopt the cosmological parameters
$(\Omega_{\rm{M}},~\Omega_{\rm{\Lambda}},~h) = (0.3,~0.7,~0.7)$.  The
stellar masses used in this Letter are calculated for a ``diet''
\citet{salpeter55} IMF \citep{bell03}.

\begin{figure}
\epsscale{1.2}
\plotone{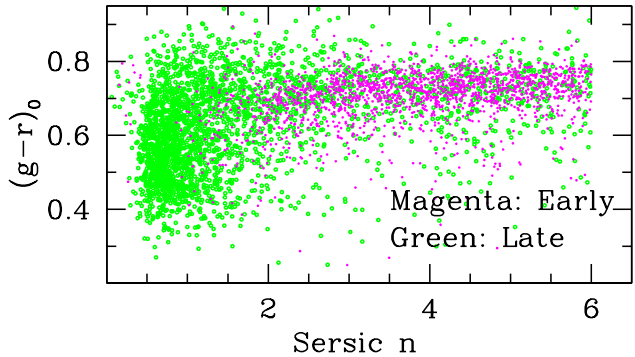}
\caption{ Rest-frame $g-r$ color vs. S\'ersic index $n$.  Magenta
 points are early-type (E+S0) galaxies, and green circles are
 late-type (Sa+later) galaxies, both according to the $B$-$n$
 method. There are significant number of late-type galaxies with high
 S\'ersic indexes.}
\label{n_col}
\end{figure}

\begin{figure}
\epsscale{1.2}
\plotone{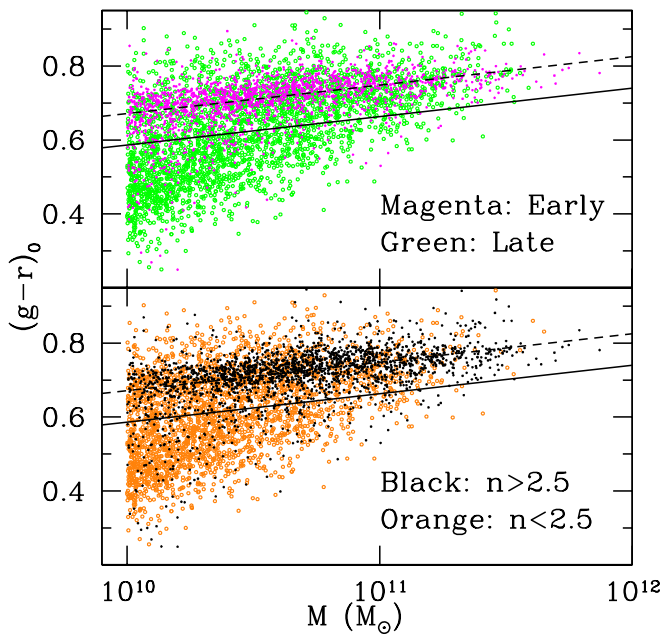}
\caption{ Rest-frame $g-r$ color vs. stellar mass. \textit{Top:}
 Magenta points are early-type galaxies, and green circles indicate
 late-type galaxies, both according to the $B$-$n$ method.
 \textit{Bottom:} Black points are galaxies with $n>2.5$, and orange
 circles are galaxies with $n<2.5$. The dashed lines show the
 least-squares linear fit to the early-type galaxies, i.e., the
 magenta points in the top panel, iteratively rejecting $3\sigma$
 outliers. The solid lines are the same as the dashed lines but
 shifted blue-ward by $2\sigma$.}
\label{M_gr}
\end{figure}

\section{DATA}\label{sec:data}
We extract a sample of galaxies from the SDSS DR5 at redshifts
$0.02<z<0.03$. This is the sample described in vdW07, but extended
down to a stellar mass of $M=10^{10}~M_{\odot}$ and limited to
$z<0.03$.  The faintest galaxies in this sample have total magnitudes
$r\sim 16.5$ or $g\sim 17.5$.  such that spectroscopic completeness is
ensured. For details concerning the determination of stellar masses,
morphologies, and environment, see vdW07. In short, stellar masses are
derived from rest-frame $g-r$ colors, using the empirical calibrations
as described by \citet{bell03}. Local surface densities are estimated
by measuring the distance to the seven-th nearest neighbor with the same
radial velocity within 1000 km s$^{-1}$. The photometric parameters
(luminosity and rest-frame color) are derived from the SDSS pipeline
archive. The $K$-corrections, to obtain the rest-frame $z=0$ colors
$(u-g)_0$ and $(g-r)_0$, and a correction on the total magnitude to
account for a known problem with the SDSS pipeline for bright
galaxies, are described in vdW07.

Morphologies are determined by measuring $n$ with GALFIT
\citep{peng02} and subsequently the dimensionless parameter $B$, which
is the rms in the residual divided by the mean of the fit within two
effective radii \citep{blakeslee06}. To classify galaxies we use the
following combination of $B$ and $n$: galaxies with $B<0.065(n+0.85)$
are considered early types (E+S0), all others late types (Sa and
later).  This criterion agrees with visually determined morphologies
90\% of the time, and, more importantly for our purposes, the
systematic difference in the relative numbers of early- and late-type
galaxies is less than 1\% \citep[see][and vdW07]{blakeslee06}.
Simulations in which we add noise to the images establish that, with
the resolution and depth of SDSS $g$-band imaging, visually determined
morphologies can be reliably reproduced with the $B$-$n$ method down
to $g=17.5$, all the way down to the faintest galaxies in our sample.

\begin{figure}
\epsscale{1.2}
\plotone{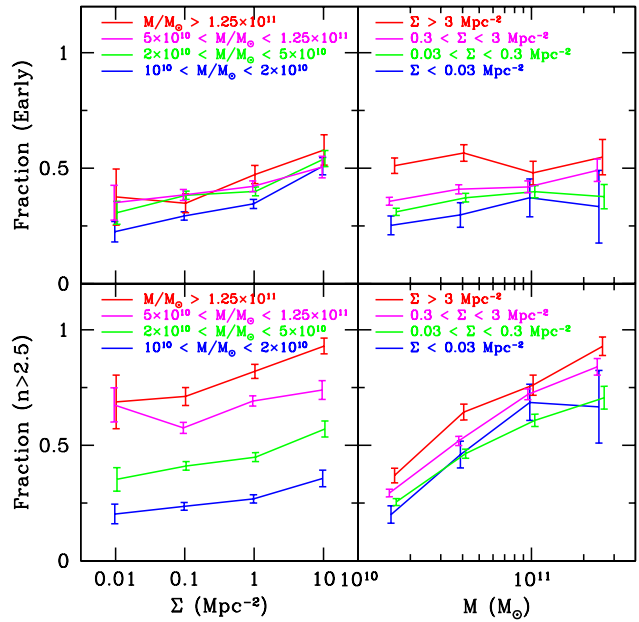}
\caption{ \textit{Top-left:} The relation between morphology and
 environment for galaxies in four different mass bins.
 \textit{Top-right:} The relation between morphology and galaxy mass
 for different density bins. \textit{Bottom-left:} The relation
 between structure (S\'ersic parameter $n$) and local surface density
 for galaxies in different mass bins. \textit{Bottom-right:} The
 relation between structure and galaxy mass in different density bins.
 Morphology depends mostly on environment, not on galaxy mass.
 Structure, on the other, hand depends mainly on galaxy mass, and only
 weakly on environment.}
\label{struct}
\end{figure}

\begin{figure}
\epsscale{1.2}
\plotone{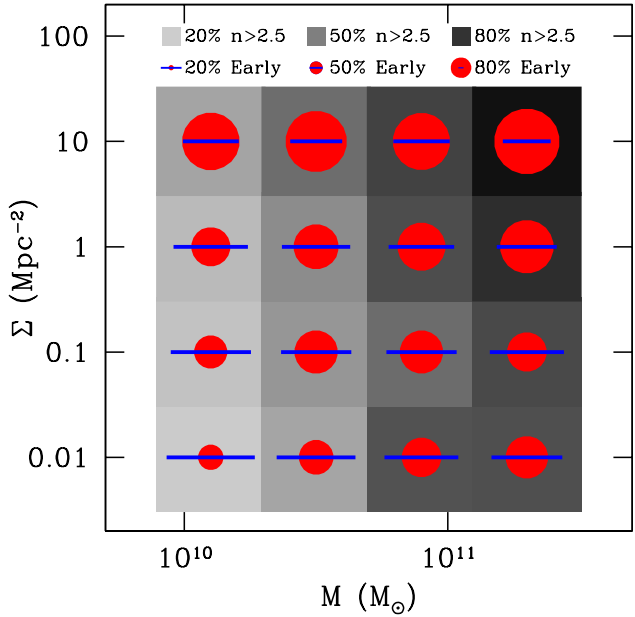}
\caption{Graphical illustration of the result that morphology mainly
 depends on environment, and structure mainly on mass.  The size of
 the filled, red circles (increasing from bottom to top) corresponds
 to the early-type galaxy fraction, the gray scale of the background
 (increasing from left to right) corresponds to the fraction of
 galaxies with $n>2.5$. }
\label{MD}
\end{figure}

\section{MORPHOLOGY AND STRUCTURE}\label{sec:morphstruct}

In the literature different observables are used to distinguish
different types of galaxies.  One can think of morphology, S\'ersic
index or concentration, and color. In Fig. \ref{n_col} we show that,
to first order, there is good correspondence between morphology
(according to the $B$-$n$ method, see Sec. \ref{sec:data}), structure
(S\'ersic index), and $(g-r)_0$ color, and that the distributions are
bi-modal for all three.  There are red, early-type galaxies with high
S\'ersic indexes, and blue, late-type galaxies with low S\'ersic
indexes.  However, there are important deviations from this simple
picture. For example, there are many red galaxies with low S\'ersic
indexes, and there are many late-type galaxies with high S\'ersic
indexes.

In order to quantify the degree of overlap between these galaxy
properties, we use three criteria, each of which separates our sample
into two sub-samples.  The first is the morphological classification
criterion based on the $B$-$n$ method. The second criterion is based
on the S\'ersic index alone (i.e., structure): galaxies with $n>2.5$
are separated from those with $n<2.5$. The third criterion is based on
color: galaxies with $(g-r)_0>0.077(\log(M/M_{\odot})-11)+0.663$ are
considered red, all others blue.  This color criterion is based on the
least-squares linear fit performed on the early-type galaxies alone
(as selected by the $B$-$n$ method), and iteratively rejecting
$3\sigma$ outliers, where $\sigma$ is the scatter in $(g-r)_0$ around
the fit (see Fig. \ref{M_gr}).  We separate blue and red galaxies by
shifting the least-squares linear fit down by $2\sigma=0.085$ mag.

Down to our mass limit of $10^{10}~M_{\odot}$ only 51\% of the red
galaxies have early-type morphologies. The red fraction among
early-type galaxies, on the other hand, is as high as 90\%. Blue,
early-type galaxies are rare, and the 10\% found in our sample could
be entirely due to misclassifications, as the random error in the
$B$-$n$ method is 10\% for individual galaxies (see vdW07).

As many as 34\% of the galaxies with $n>2.5$ have late-type
morphologies, which may come as a surprise, since usually galaxies
with high S\'ersic indexes, i.e., de Vaucouleurs profiles, are thought
of as bulge-dominated early types. An important aspect of this
difference is illustrated in Fig. \ref{M_gr}: the high-$n$ galaxies
(\textit{black data points in the bottom panel}) have systematically
higher masses than the early-type galaxies (\textit{magenta data
points in the top panel}).  Apparently, the mass distribution of
galaxies with high S\'ersic indexes is different from the mass
distribution of early-type galaxies.

The different behavior of morphology and structure as a function of
mass is illustrated further in the right-hand panels of
Fig. \ref{struct}.  The bottom right panel shows that the fraction of
galaxies with $n>2.5$ depends strongly on galaxy mass. On the other
hand, the early-type fraction does not change significantly with
increasing mass.

The strong correlation between structure and mass is a reproduction of
the result by \citet{kauffmann04} even though those authors measure
structure differently, by the concentration index $C$, the ratio
between the radii containing 90\% and 50\% of the light in the
$r$-band.  This similarity is no surprise as $C$ and $n$ both depend
solely on the change in the slope of the surface brightness profile
with radius.

Structure and morphology not only behave differently with respect to
galaxy mass, but also with respect to environment: for galaxies with a
given mass, the fraction of early types increases with projected
surface density (top left panel of Fig. \ref{struct}). The dependence
of morphology on environment is therefore stronger than the dependence
of morphology on mass.  For structure this trend is reversed: the
fraction of galaxies with $n>2.5$ also modestly increases over the
same range in density; however, the dependence of galaxy mass is much
stronger (bottom-left panel of Fig. \ref {struct}).  The remarkable
inter-dependencies of morphology, structure, mass, and environment are
visually illustrated in Fig. \ref{MD}.

It is noteworthy that the dependence of galaxy structure on mass is
much stronger than the morphological dependence on environment.  Over
3 orders of magnitude in local surface density, early-type fractions
change only from $\sim 25\%$ to $\sim 50\%$, whereas the fraction of
galaxies with $n>2.5$ increases from $\sim 25\%$ to $\sim 75\%$ over
little more than 1 order of magnitude in mass (see Fig. \ref{struct}).
In that respect environment only plays a secondary role in shaping the
galaxy population.

\begin{figure}
\epsscale{1.2}
\plotone{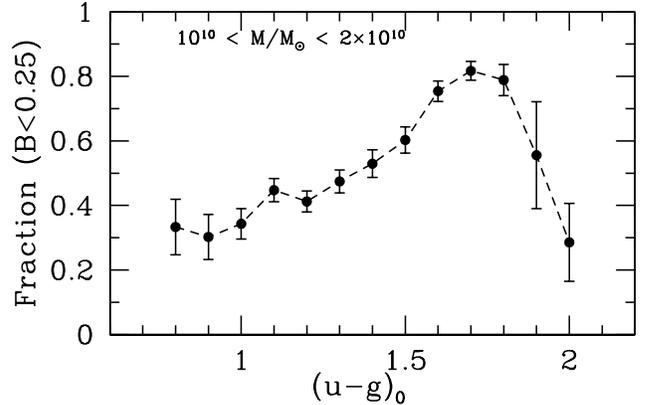}
\caption{ The fraction of smooth galaxies (with $B<0.25$) as a
 function of color for low-mass galaxies.  The fraction of smooth, red
 galaxies is much larger than the fraction of smooth, blue galaxies
 (with the notable exception of the reddest galaxies).  This implies
 that star formation is responsible for the increased bumpiness of
 blue galaxies.}
\label{colorB}
\end{figure}

It is intuitively clear that the key to understanding the difference
between structure and morphology, which is solely due to the bumpiness
parameter, is star formation activity.  The definition of morphology
as used in this Letter, i.e., a combination of $n$ and $B$, which is a
measure of the deviation from a smooth profile, explicitly uses signs
of star formation (spiral arms and star-forming regions) to separate
late-type from early-type galaxies.  This was quantified earlier by
\citet{takamiya99}, who showed that residuals from smooth profile fits
correlate with H$\alpha$ emission lines maps.  Furthermore,
\citet{baldry06} show that color, like morphology, depends on
environment for galaxies with a given mass, trends that we also see in
our sample.

In Fig. \ref{colorB} we explicitly show the relation between color and
$B$. For low-mass galaxies, which have a large range in color, the
fraction of smooth galaxies, those with $B<0.25$, increases toward
redder colors.  A notable deviation from this trend is that the
reddest galaxies with $(u-g)_0\gtrsim 1.85$ are generally not
smooth. Visual inspection of this small number (5\%) of very red
galaxies shows that these are edge-on disk galaxies and irregular
galaxies, such that their colors can be explained by high extinction.

Since star formation is generally confined to disks, color gradients,
like bumpiness, are expected to trace morphology. \citet{park05}
design color versus color-gradient criteria that, combined with
concentration index, effectively distinguish visually classified
early- and late-type galaxies. Following up on this, \citet{park07}
find results similar to those we present here: they find that, at
fixed luminosity, morphology depends on environment, whereas the
dependence of the concentration index on environment is much
weaker. It is encouraging that these results and ours are in
agreement, because it is not self-evident that a morphological
classification method based on colors behaves the same way as a method
that involves spatial information that more directly corresponds to
visual classifications.

We note that morphology behaves intrinsically differently as a
function of stellar mass and luminosity. \cite{park07} find at fixed
density a rather strong relation between luminosity and early-type
galaxy fraction.  At luminosities for which our sample is complete, we
also see an increased early-type fraction of high-luminosity
galaxies. It is beyond the scope of this Letter to fully address this
issue.

\section{CONSEQUENCES FOR HIGH-REDSHIFT STUDIES}

At higher redshifts ($z\sim 1$) often some measure of galaxy structure
is used to separate early-type from late-type galaxies
\citep[e.g.,][]{abraham96}.  Even more advanced measures of the light
distribution than the concentration parameter $C$, such as $G$, the
Gini coefficient, and $M_{20}$, the second-order moment of the
brightest 20\% of the flux \citep{lotz04, capak07, zamojski07}, are
essentially a measure of concentration for ``normal'' galaxies. $G$
and $M_{20}$ are very suitable to distinguish star-bursting and
merging galaxies, but for early-type and spiral galaxies the
correlation between concentration and either of those parameters is
too tight to effectively select the early types.

Even in classification methods which use asymmetry in addition to
concentration \citep[e.g.,][]{schade95}, the problem sketched above is
not necessarily remedied adequately: many late types, especially the
massive ones, are rotationally symmetric, such that invoking asymmetry
as a morphological classifier does not help much in distinguishing
between Sb, Sa, and S0 galaxies \citep[see, e.g.,][]{conselice03}.
Only a parameter that quantifies the residual from a smooth light
distribution can be successful in making that distinction.  The
bumpiness $B$ used in this Letter is one example of such a parameter;
another one, the ``clumpiness'' parameter $S$, was introduced earlier
by \citet{conselice03}, although \citet{blakeslee06} note that $S$ and
$B$ are not interchangeable and that the combination of $n$ and $B$
seems to be more successful in separating early- and late-type
galaxies.

Our conclusion that structure and morphology are intrinsically
different galaxy properties has consequences for the interpretation of
high-redshift morphological studies if those are based on some measure
of structure. The obvious problem is that the fractions of highly
concentrated and early-type galaxies are different, and that a direct
comparison at different redshifts can lead to biases in the inferred
evolution. A more subtle problem arises when the evolution in
``morphological'' properties is measured for galaxies with different
masses or in a variety of environments.  As we saw in
Sec. \ref{sec:morphstruct}, morphology and structure differ in their
behavior as a function of mass and environment. Therefore, observed
changes in, e.g., concentration are easily misinterpreted in terms of
the effect of the environment or galaxy mass on the evolution of
galaxies.

This type of problem can only be avoided if the methods used to
classify low- and high-redshift galaxies are as similar as possible.
This can be achieved by consistently using the same automated methods
or by taking great care in using a fixed set of visual classification
criteria for galaxies at different redshifts \citep[as in,
e.g.,][]{postman05}.

\section{CONCLUSIONS}

We have shown that structure (in this Letter quantified by the
S\'ersic parameter $n$) and morphology (quantified by $n$ and the
bumpiness parameter $B$) are distinct galaxy properties. There is a
significant number of galaxies with high S\'ersic indexes but
late-type morphologies (Fig. \ref{n_col}). The physical significance
of the difference between structure and morphology becomes apparent
when their behavior as a function of galaxy mass and environment is
analyzed. Structure mainly depends on galaxy mass whereas morphology,
at fixed galaxy mass, depends on environment (Fig. \ref{struct}).  The
different behavior is linked to star formation activity, which only
weakly affects the structure of a galaxy, but strongly affects its
morphological appearance (Fig. \ref{colorB}). This implies that the
existence of the MDR is intrinsic, and is explained by the
environmental dependence of star formation activity. This seems
trivial; however, it means that the MDR is not simply the result of
underlying correlations, i.e., the strong dependence of structure and
star formation on galaxy mass, and the environmental dependence of the
mass function.

The author gratefully acknowledges financial support from NASA grant
NAG5-7697, and is indebted to support from John Blakeslee and Holland
Ford. Stimulating discussions with Andrew Zirm and Marijn Franx helped
shape this Letter. Attentive reading by and valuable comments from
Andrew Zirm, Roderik Overzier, Alison Crocker, John Blakeslee,
Bradford Holden, Ricardo Demarco, and Marc Postman have greatly
improved the readability.

\bibliographystyle{apj}


\end{document}